\documentstyle[preprint,eqsecnum,aps,prb,amstex]{revtex}

\begin{document}
\title{Staircase effect in metamagnetic transitions of charge and orbitally ordered
manganites.}
\author{V. Hardy, S. H\'{e}bert, A. Maignan, C. Martin, M. Hervieu and B. Raveau.}
\address{Laboratoire CRISMAT, Unit\'{e} Mixte de Recherches 6508, Institut des\\
Sciences de la Mati\`{e}re et du Rayonnement - Universit\'{e} de Caen, 6\\
Boulevard du\\
Mar\'{e}chal Juin, 14050 Caen Cedex, France.}
\date{\today}
\maketitle

\begin{abstract}
This paper reports on peculiar metamagnetic transitions which take place in
antiferromagnetic, charge and orbitally ordered (AFCOO) manganites. At very
low temperatures, the virgin magnetization curves of some of these compounds
exhibit several, sharp steps giving rise to a staircase-like shape. This
staircase effect is shown to be sensitive to various experimental
parameters. Several potential interpretations of the staircase effect are
discussed in relation to this set of results. A martensitic-like scenario,
involving a leading role of the structural distortions associated to the
collapse of the orbital ordering, is found to be the most plausible
interpretation.
\end{abstract}

\pacs{}

\section{Introduction}

Mixed-valent manganites of formulation RE$_{1-x}$AE$_x$MnO$_3$ (RE being a
trivalent rare-earth ion, and AE a divalent alkaline-earth ion) exhibit a
great variety of fascinating properties, such as charge ordering (real-space
ordering of the Mn$^{3+}$and Mn$^{4+}$species) and colossal
magnetoresistance (collapse of the resistivity by several orders of
magnitude under application of an external magnetic field).\cite{REVIEW}
Charge ordering is often accompanied by a long-range ordering of the Mn$%
^{3+} $ e$_g$ orbitals, leading to prominent structural distortions. In Pr$%
_{1-x}$Ca$_x$MnO$_3$ compounds with $x\approx 0.5$, the small A-site
cationic radius and the Mn$^{4+}/$Mn$^{3+}$ ratio close to 1 contribute to
the setting up of a particular groundstate, which is both charge and
orbitally ordered, insulating and antiferromagnetic of CE type (hereafter
referred to as AFCOO). Application of magnetic fields can affect the subtle
energy balance in these systems, and trigger the development of a
ferromagnetic, conducting state (hereafter referred to as F). This process
is at the origin of the most spectacular effects of colossal
magnetoresistance in manganites. However, very high magnetic fields are
required to melt the robust AFCOO state present for $x=0.5$ (about 25 T at 4
K).\cite{TOK98} For compositions shifted towards $x=0.3$, lower fields are
needed (e.g., about 5 T for $x\approx 0.35$).\cite{LEE95} Mn-site
substitutions can also weaken the AFCOO state,\cite{RAV97} making it
susceptible to undergo an AFCOO-F transition at low-T under moderate fields.

Recently, we observed very intriguing metamagnetic transitions in such Pr$%
_{0.5}$Ca$_{0.5}$Mn$_{1-y}$M$_y$O$_3$ compounds, M being a magnetic or
nonmagnetic cation.\cite{STEP1,STEP2} In some cases, the virgin
magnetization curve at low-T indeed displays successive, abrupt steps as the
field is increased. This staircase-like shape of the $M(H)$ curves
disappears as $T$ is increased above about 10 K. Similar features were
observed in resistivity and specific heat measurements.\cite{STEP1}

In this paper, we investigated by isothermal magnetization loops several
aspects of this staircase effect : sample dependence, role of temperature,
cycling effects, mode of field variation, role of microstructure, type of
disorder affecting the AFCOO stability. Most of the experiments were carried
out on a set of samples of the composition Pr$_{0.5}$Ca$_{0.5}$Mn$_{0.95}$Ga$%
_{0.05}$O$_3$. Complementary measurements were also performed at very low
temperatures in Pr$_{1-x}$Ca$_x$MnO$_3$ compounds with $0.35\leq x\leq 0.40$.

\section{Experimental details}

All the samples used in this study are ceramics, except one crystal which
was grown in a mirror furnace. The Pr$_{0.5}$Ca$_{0.5}$Mn$_{0.95}$Ga$_{0.05}$%
O$_3$ and Pr$_{1-x}$Ca$_x$MnO$_3$ ceramic samples were prepared by
conventional solid-state reaction. Stoichiometric mixtures of the oxides Pr$%
_6$O$_{11}$, CaO, MnO$_2$ and Ga$_2$O$_3$ were intimately ground and the
powders, first heated at 1000 ${{}^{\circ }}$C, were pressed in the form of
bars. The sintering was made at 1200${{}^{\circ }}$C and at 1500${{}^{\circ }%
}$C for 12 h, then the bars were slowly cooled to 800${{}^{\circ }}$C at 5${%
{}^{\circ }}$C/h, before being quenched to room temperature. A Pr$_{0.63}$Ca$%
_{0.37}$MnO$_3$ crystal was cut out of the central part of a several-cm-long
specimen grown in a mirror furnace using a floating-zone method. Physical
characterizations of this crystal were previously reported.\cite{0604n1} The
purity of all samples was attested to by X-ray powder diffraction and
electron diffraction studies. Energy-dispersive X-ray microanalyses yielded
cationic compositions in agreement with the nominal ones, within the
accuracy of this technique. Electron diffraction investigations also
revealed the existence of twinning domains in all samples, including the
single crystal.

The magnetic measurements were carried out by means of a Quantum Design
extraction magnetometer, with fields up to 9 T and temperatures down to 2.5
K. All hysteresis loops were recorded after a zero-field cooling from room
temperature. It must be noted that room temperature (RT) is larger than all
transition temperatures -corresponding to spin, charge or orbital orderings-
that can be encountered in these materials. Even for the unsubstituted,
half-doped Pr$_{0.5}$Ca$_{0.5}$MnO$_3$, the charge and orbital ordering
takes place at $T_{CO}\simeq 250$ K, well below RT. The hysteresis loops
were registered according to the following procedure: The magnetic field was
increased from 0 to 9 T, and then decreased down to 0 T, with equally spaced
values (0.25 T in most cases) ; a waiting time is imposed between the end of
the field installation and the beginning of the measurements (1 min in most
cases); then, three measurements are successively recorded (all shown in the
figures). Complementary experiments were carried out to investigate the
influence of field spacing and waiting time values.

For Pr$_{0.5}$Ca$_{0.5}$Mn$_{0.95}$Ga$_{0.05}$O$_3$, that is the compound
used for most of the experiments, we studied four ceramic samples coming
from the same batch, and having nearly the same mass and shape. The first
measurement recorded on each of these four samples -hereafter labelled as A,
B, C and D- was an hysteresis loop at 5 K. This allowed us to properly
address the question of reproducibility, by considering only samples being
in a {\it pure} virgin state, i.e., without any previous application of high
fields at low temperature.

\section{Results}

Figure 1 shows hysteresis loops, recorded at 5 K, in Pr$_{0.5}$Ca$_{0.5}$Mn$%
_{0.95}$Ga$_{0.05}$O$_3$ and Pr$_{0.5}$Ca$_{0.5}$MnO$_3$ [hereafter denoted
as (PrCa50)Ga5\% and PrCa50, respectively]. In PrCa50, one observes a
linear, reversible curve up to 9 T, consistent with the fact that this
compound stays in a pure AFCOO state within this field range. In
(PrCa50)Ga5\%, the field-increasing branch of the loop also exhibits a
linear behavior for magnetic fields lower than about 4 T. However, the slope
is significantly larger than in PrCa50. This was confirmed by low-field
ac-susceptibility curves, which also revealed the existence of a cusp around
50 K in the Mn-site substituted compounds.\cite{STEP2} Such a cusp suggests
that the introduced disorder on the Mn-sublattice allows the development of
competing ferromagnetic interactions within the antiferromagnetic matrix.
These results point to the setting of a very weakened AFCOO groundstate in
(PrCa50)Ga5\% as compared to PrCa50. For fields larger than 4 T, the curve
of (PrCa50)Ga5\% (sample A) exhibits three large jumps, separated by almost
flat plateaus. Approaching 9 T, the magnetization undergoes a small increase
from the last plateau, leading to a value of 3.25 $\mu $$_B$ $/$ f.u.. Since
the expected magnetization for full polarization of the Mn spins in
(PrCa50)Ga5\% is 3.3 $\mu $$_B$ $/$ f.u.,\cite{Rq1} the sample under 9 T is
virtually in a saturated magnetic state. As the field is decreased, one
observes a rather flat curve down to about 2 T, followed by a rapid drop of
magnetization yielding a remanent value of about 0.15 $\mu $$_B$ $/$ f.u..
The inset of Fig. 1 displays following field-scans recorded on this sample,
first by extending the field decrease down to -9 T, and then by
re-increasing it up to 9 T. The curve between 0 and -9 T is symmetrical to
the reverse leg of the first quadrant. Furthermore, the curve recorded when
increasing the field from -9 T to 9 T is almost superimposed on the one
obtained when decreasing from 9 T to -9 T. After the first application of 9
T, the sample thus seems to behave as a bulk, reversible ferromagnet.

Before going further, it is essential to check the reproducibility of the
staircase-like behavior. Hysteresis loops at 5 K were recorded on four
virgin samples belonging to the same batch. For the sake of clarity, figure
2 just shows the field-increasing parts of the loops. First, one can point
out the remarkable superimposition of the linear regimes found between 0 and
4 T, attesting to the good homogeneity of the cationic composition within
this set of samples. For each of them, three large steps occur within the
field range 4-9 T, and the magnetization reaches about 3.25 $\mu $$_B$ $/$
f.u. under 9 T in all cases. However, it is clear that the characteristic
fields associated with the steps slightly differ from sample to sample. A
significant scatter exists also about the magnetization values corresponding
to each plateau, especially for the first one. These results demonstrate
that the staircase effect only exhibits a semi-quantitative reproducibility.

Figure 3 displays hysteresis loops recorded on sample B at 5, 7.5 and 10 K.
All curves have nearly the same initial slope, saturation value and remanent
magnetization, but the steps have completely disappeared for $T=10$ K. At
this temperature, the metamagnetic transition rather has a smooth, $S$-shape
centered around 5 T. For $T\leq 5$ K, the transition exhibits several steps
(observed down to our lowest temperature $T=2.5$ K). The curve at 7.5 K well
illustrates the crossover in temperature. As compared to lower temperatures,
the curve is smoothed out from both sides of the field range, leaving just
one step in the middle of the transition. This behavior is basically
different from a simple broadening of the steps seen at 5 K. One can also
notice that the curves with steps seem to ``zigzag'' around the one at 10 K,
a behavior suggesting that this last curve might be representative of an
underlying equilibrium $M(H)$ curve at low-T.

In our first report of the staircase effect,\cite{STEP1} we already
mentioned the existence of some variations in the pattern of steps as the
sample is cycled between low-$T$ and RT. To get a direct insight into this
point, we have successively recorded on sample C five hysteresis loops at 5
K, each of them starting from a zero-field cooled state. The results are
shown in Fig. 4. For the first three loops, the step structure is well
reproducible. One just observes small variations in the height of the second
plateau and in the third step field ($H_{S3}$). In contrast, a marked
evolution takes place in between the third and fourth runs: One observes
clear upward shifts in the step fields $H_{S1}$ and $H_{S2}$, while the
third step is replaced by a smooth tail. Then, the fifth loop remains well
superimposed on the preceding one. Along these five runs, the initial slope
and remanent magnetization remain constant, whereas there is a systematic,
small decrease of the maximum magnetization. All these features have been
observed in other samples, except that ``switches'', such as that seen in
Fig. 4 between runs 3 and 4, can occur sooner or later.

All previously shown data were recorded by using a field spacing ($f_s$)
equal to 2500 Oe and a waiting time ($t_w$) equal to 1 min. Figure 5 shows
three loops successively recorded on sample C at $T=5$ K, following
different procedures: (1) $f_s=2500$ Oe with $t_w=1$ min (run 5 of Fig. 4);
(2) $f_s=1000$ Oe with $t_w=1$ min; (3) $f_s=2500$ Oe with $t_w=5$ min. The
last two runs lead to the same average sweep rate (35 mT / min), including
field installation, waiting time and data acquisition. For a constant
waiting time (1 min), one observes that using smaller field spacings pushes
the steps to higher fields. These shifts in the $H_S$ values are accompanied
by upward shifts of the corresponding plateaus. Run 3 registered with $%
f_s=2500$ Oe and $t_w=5$ min yields a curve nearly superimposed on that of
run 1 obtained for $fs=2500$ Oe and $t_w=1$ min. Thus one can conclude that:
(i) the shifts in $H_S$ observed between runs 1 and 2 can actually be
ascribed to the variation in $f_s$, and not just to a cycling effect (as in
Fig. 4); (ii) the impacts of increasing the average sweep rate via $f_s$ or $%
t_w$ are completely different. It turns out that the field spacing is
actually the relevant parameter at the origin of the difference between run
1 and 2.

Up to now, the staircase effect has only been observed in Mn-site
substituted PrCa50 samples. One can wonder whether this property is related
to the weakening of the AFCOO state, or more specifically to a disordering
of the Mn sublattice. Thus, we re-investigated Pr$_{1-x}$Ca$_x$MnO$_3$
compounds with $x$ values yielding AFCOO states able to be destabilized
under moderate fields. We studied two ceramic samples with $x=0.40$ and $%
x=0.35$, as well as a crystal with $x=0.37$. For all of them, smooth
metamagnetic transitions were observed for temperatures down to 5 K. At 2.5
K, in contrast, the hysteresis loops display very steep transitions, as
shown in Fig. 6. The ceramics with $x=0.35$ and $x=0.40$ present a unique,
large step at the transition, while a multisteps structure takes place for
the crystal with $x=0.37$. This staircase effect is exactly the same as that
encountered in Mn-site substituted compounds with $x=0.50$. This is an
important piece of information which demonstrates that the multisteps
behavior is not basically related to the disordering of the Mn-sublattice
induced by magnetic or nonmagnetic foreign cations. In other respects, one
can wonder why the PrCa37 crystal behaves differently than the ceramics
having bordering compositions. These issues will be addressed below.

\section{Discussion}

Let us try to consider the various, possible interpretations of the
staircase-like metamagnetic transitions presently observed in
antiferromagnetic, charge and orbitally ordered (AFCOO) manganites.

First, one must notice that multisteps metamagnetic transitions are not rare
in the literature on antiferromagnets.\cite{STR77} This is connected to the
existence of a ``ferrimagnetic'' domain at low-$T$ in the phase diagram,
containing one or several intermediate magnetic arrangements in between the
antiferromagnetic and paramagnetic phases. In such cases, it must be
emphasized that the step fields are true {\it critical fields} given by
combinations of different exchange fields present in the system. Therefore,
they must be well reproducible and independent of the measuring procedure,
in contrast to the behaviors found in AFCOO manganites. In addition, the
magnetization of each plateau should also have well defined values, a
feature which again strongly departs from our observations.

The literature provides us with another series of results that could be
relevant to our situation. In site-diluted antiferromagnets, it was indeed
shown that metamagnetic transitions can display several abrupt steps. This
was well illustrated in the case of Fe$_{1-x}$Mg$_x$Cl$_2$ for which
experimental and simulated $M(H)$ curves bear close resemblance to ours.\cite
{KUS96} This behavior was interpreted in terms of avalanches of spin
reversals in the AF structure. Random distribution of magnetic vacancies (Mg$%
^{2+}$ ions) generates a great variety of local situations in the spin
lattice. Different values of external field are required to flip a spin in
these various neighborhoods. As it is increased, the field first reaches a
value inducing spin reversals within the most disturbed regions, probably
close to clusters of magnetic vacancies. Once a first spin has been flipped,
it modifies the environment of its neighbors which can themselves flip, and
so on. This flipping process can propagate as a {\it domino-like wave}
within the spin structure,\cite{MAR01} resulting in sizeable avalanches.
Simulations showed that these avalanches can be blocked by regions locally
free of defects or by clusters of previously flipped spins.\cite{KUS96}
Then, the field can be further increased over a wide range without
triggering new spin reversal, leading to almost constant magnetization. This
process yields $M(H)$ curves with large, sharp steps connected by plateaus,
just like in our case. Nevertheless, some experimental results cast doubts
on the relevancy of this mechanism to our situation. Firstly, we have found
that the staircase effect also exists in Pr$_{0.63}$Ca$_{0.37}$MnO$_3$,
i.e., without substitution on the spin lattice itself. Secondly, the
mechanism described above involves true critical fields, each of them being
associated to a particular local arrangement (number and spatial
distribution of adjacent vacancies).\cite{KUS96} These critical fields are
expected to be reproducible for a given sample. Once again, the
modifications in the pattern of steps resulting from cycling or varying
procedure cannot be accounted for in the frame of such a scenario.

Actually, the whole set of data indicates that the staircase effect is
rather dealing with a succession of metastable states (corresponding to the
plateaus), separated by boundaries (the step fields) which values cannot be
perfectly reproducible. Keeping this in mind, let us turn back to the
specific case of our AFCOO manganites. In such compounds, the metamagnetic
transition is widely recognized to proceed through nucleation and growth of
F domains within the AF matrix. The most natural idea about the occurrence
of multisteps structures consists in invoking additional mechanisms which
could impede the development of these nucleation/growth processes. When such
a competition involves opposing trends which are very different in nature,
it can yield a discontinuous response in various physical systems. For
instance, it is the case in ferromagnets where Barkhausen noise results from
jerky motions of domain walls, a behavior often attributed to pinning by
defects. Related phenomena are also encountered in martensitic
transformations (the so-called Acoustic Emission effect). In manganites, it
turns out that transitions involving the COO state could have a
martensitic-like character. This analogy was previously pointed out in
recent publications.\cite{UEH00,MAH01,POD01} A martensitic transformation
(MT) is basically a first-order transition (FOT) involving a shear-induced
lattice distortion between a high-$T$ phase (austenite) and a low-$T$ phase
(martensite). Because of a significant change in shape of the unit cell, the
nucleation of first martensitic domains upon cooling induces long-range,
anisotropic strains developing from the martensite/austenite interfaces.\cite
{VIV94a} The resulting increase of elastic energy can transitorily block
further growth of martensitic phase under continuous lowering of temperature
(the external parameter driving the FOT in this case). Significant
undercooling is actually required to make the transition starting again.
This sequential process repeats itself along the course of the
transformation. Finally, despite its intrinsic first-order character, the MT
is spread over a large range of temperature between $M_S$ and $M_F$ (start
and finish temperatures of the austenite to martensite transformation,
respectively).

The settling of COO in manganites is well known to result in significant
distortions of the unit cell. This originates from cooperative Jahn-Teller
effects, giving rise to long-range ordering of the Mn$^{3+}$ e$_g$ orbitals.
It must be noted that such phenomena related to internal strains can be in
play during either the setting or the collapse of COO domains. Furthermore,
similar effects can be expected when COO domains coexist with either
paramagnetic or ferromagnetic (F) phases. The previous studies invoking the
martensitic analogy referred to temperature-driven transitions between the
paramagnetic and COO phases at $T_{COO}$.\cite{UEH00,POD01} In the present
paper, we rather consider the AFCOO-F transition, driven by application of
magnetic field at low-$T$. The two phases involved in this transition have
very different cell parameters, which can justify a martensitic-like
character of the transformation. In order to quantify this point, let us
compare the results of neutron diffraction data recorded at 10 K under zero
field (3T2 at LLB, Saclay, France) inPrCa50 (a good reference for the AFCOO
phase) and (PrCa50)Cr5\% (which is ferromagnetic at low-$T$ even under zero
field).\cite{DAM98,JIR00} When going from the AFCOO (PrCa50) to the F phase
(PrCa50Cr5\%), $a$ and $c$ parameters ($Pnma$ space group) are decreased by
0.6\% and 0.8\%, respectively, while $b$ is increased by about 1.7\%. Since
the cell parameters of the two components are not strictly equal at 300 K,
it is still more reliable to compare a distortion parameter, such as $%
D=(100\,/\,3)\sum_{i=1,2,3}\left( \left| a_i-\left\langle a\right\rangle
\right| /\left\langle a\right\rangle \right) $, where a$_i$ refers to the
three cell parameters, and $\left\langle a\right\rangle $ is the average
value. While the $D$ values of both compounds are similar in the
paramagnetic phase at 300 K [0.11 and 0.17 for (PrCa50)Cr5\% and PrCa50,
respectively], these values are totally different in the ordered phases at
10 K: $D=0.15$ in (PrCa50)Cr5\%, and $D=1.19$ in PrCa50. These structural
evolutions clearly indicate that large, anisotropic strains must develop at
the F/COO interfaces during the metamagnetic transformation. In a canonical
MT, the austenite phase transforms into the martensite upon lowering
temperature or applying external stress. In our case, the roles of austenite
and martensite are played by the COO and F phases, respectively, while
increasing magnetic field just acts as lowering temperature or increasing
stress. Contrary to the case of standard MT, note that the most distorted
phase is the ``austenite'' in this analogy.

Let us now try to describe the occurrence of multisteps on $M(H)$ curves in
the frame of such a martensitic scenario. At low-$T$ and under zero-field,
the COO (austenite) phase is extremely predominant, as attested to by the
linearity of the $M(H$) curves. For sufficiently high magnetic fields, F
domains (martensite) start growing. A slight deviation from linearity can be
detected on the curves above about 3.5 T. This upward curvature, however, is
very small, which can be ascribed to the inhibiting effect of the strains at
the F/COO interfaces. As the field is continuously increased, the driving
force acting on the spins increases. There is a field value for which this
force becomes high enough to locally overcome the elastic constraints,
triggering a sudden motion of F/COO interfaces in a part of the sample.
This, in turn, destabilizes the local stress field and, in an avalanche-like
way, it can result in spreading the F state over large regions of the
sample. Such an abrupt increase of the martensite volume fraction (F phase
in our case) yields a sizeable increase of magnetization. Along this
avalanche process, the magnetic energy decreases while the elastic one
increases, a balance which can lead the system to be frozen in another
metastable state. The overall transition may thus proceed by successive
jumps between metastable states, which directly yield a staircaselike shape
on $M(H)$ curves. One could argue that the discontinuous character of the
transition in most MT takes place at small scales, making it hardly visible
on the evolution of bulk quantities. However, there are some cases for which
MT develops by successive ``bursts''. For instance, the volume fraction of
martensite versus temperature in Ni-Ti alloys can increase by macroscopic
steps,\cite{SKR91} a behavior consistent with the existence of the staircase
effect seen in $M(H)$ curves of our manganites.

Let us now test more precisely the consistency of this martensitic-like
scenario with the whole set of experimental results. Martensitic
transformations are known to be basically sensitive to defects, which can
influence the nucleation and growth of the product phase within the parent
one (pre-existing defects or defects induced by previous transformations).
Such a role of defects can explain some slight variations from sample to
sample as shown in Fig. 2. The evolution observed between successive $M(H)$
curves can also be ascribed to the influence of defects created by the
lattice distortions along the COO-F transformations. In MT, prominent
cycling effects are observed when scanning the transition either by
temperature or stress variations. In most cases, it was shown that defects
produced by thermal cycling act as obstacles making the martensitic
transformation more and more difficult.\cite{HUM91} This leads to a decrease
of $M_S$, which is qualitatively consistent with the increase of the step
fields shown in Fig. 4. Figure 5 shows that the field spacing can affect the
location of the steps. As a general rule, the collapse of any {\it brittle}
metastable state can be expected to be slightly delayed by smoothly varying
the driving (destabilizing) parameter (i.e., the field in our case). The
results of Fig. 5 are qualitatively consistent with such a behavior since
one observes that smaller field increments tend to push up the breakdown
fields (just as in sandpile experiments, when adding grains of sand one by
one instead of by packs).

Another important result one has to face is displayed in Fig. 6. Why does
one observe a staircase effect in the PrCa37 crystal and not in the PrCa35
and PrCa40 ceramics ? Is it related to the $x$ value or to the sample nature
? As soon as $x$ departs from 0.5, there must be some disorder in the charge
and orbital ordering. In disordered systems for which an underlying
first-order transition proceeds through avalanches, the degree of disorder
was found to critically influence the development of the transition.\cite
{VIV94b} For low levels of disorder, the transition takes place as a single
infinite avalanche, while for high levels of disorder, it has a smooth shape
made of tiny avalanches (undetectable on the hysteresis loop of a magnetic
system, for instance). In between these situations, there is a critical
value of disorder for which the transition proceeds through superimposition
of avalanches of all sizes. One might thus speculate that some levels of
disorder just below the critical value could lead to the appearance of
several macroscopic steps on $M(H)$ curves. Nevertheless, this
interpretation is made unlikely by the fact that both PrCa35 and PrCa40
exhibit similar shapes of $M(H)$, that are consistent with low disorders.
Actually, it seems more plausible to rather invoke the difference of
microstructure between crystals and ceramics. Our ceramics are made of
grains of typical size close to 100 $\mu $m. A role of the grain size was
previously reported in the development of martensitic-like transitions in
manganites. Upon cooling across $T_{COO}$, it was found that large grains
are more favorable to the establishment of the COO phase.\cite{LEV00,POD01}
This was ascribed to a better accommodation ability of the martensitic
strains as the grain size increases. In our case, the results of Fig. 5
rather suggest that accommodation is more difficult in the crystal than in
the ceramics. The literature of MT in metallic alloys shows that there are
antagonist effects associated to grain size, leading to various global
behaviors. There are cases where the width of the martensitic transition ($%
M_S-M_F$) was found to increase with the grain size.\cite{SAK95} This
suggests that the hindering effects associated to internal stresses can
increase with size, a feature which might be consistent with our results.
The influence of microstructure on the staircase effect in manganites
obviously deserves further investigations.

\section{Conclusion}

The staircase effect recently discovered in some antiferromagnetic, charge
and orbitally ordered (AFCOO) manganites, has been extensively studied by
magnetic measurements (isothermal $M$ versus $H$ loops). Various parameters
were investigated: Reproducibility from sample to sample; influence of the
temperature; cycling effects; measuring procedure. Experimental results
allowed us to discard some of the interpretations which could be a priori
considered. An important point is that the step fields cannot be regarded as
``pure critical fields''. These characteristic fields rather separate
metastable states, and their precise values can change with many parameters.
It is also determinant to point out that the AFCOO to F transformation
involves large structural distortions.\ Actually, there are striking
analogies with the martensitic transformations encountered in metallic
alloys. We found that all our experimental features can be reasonably well
accounted for within such a scenario.

Although a staircase effect was first evidenced in Mn-site substituted
compounds, the exact role of defects in this phenomenon appears to be quite
complex. Firstly, we found that this effect can also take place in ``pure''
Mn compounds. Secondly, our martensitic interpretation is based on
structural distortions which also take place in the absence of any disorder.
The facts remain that the presence of disorder can be determinant since it
favors the creation of deep metastable states separated by large barriers in
the energy landscape, a situation well suited to the development of large
avalanches.\cite{VIV94b}

A lot of aspects of the staircase effect will have to be further
investigated. For instance, one will have to check whether it could also be
observed in the metamagnetic transitions of other kinds of AF groundstates
encountered in manganites, such as the A- or C-types. Relaxation experiments
(under fixed $T$ and $H$) should also be carried out to investigate further
the role of temperature, addressing in particular the question of the
athermal\cite{PER01} (or not) character of these multisteps transitions.

\section{Acknowledgments}

The authors thank L. Herv\'{e} (CRISMAT, Caen) for the crystal growth, and
R. Mahendiran for useful discussions about martensitic transformations.

\section{Figure Captions}

\begin{description}
\item[FIG. 1.]  Magnetization loops at 5 K in Pr$_{0.5}$Ca$_{0.5}$Mn$_{0.95}$%
Ga$_{0.05}$O$_3$ (circles) and Pr$_{0.5}$Ca$_{0.5}$MnO$_3$ (squares). Inset
displays complete scans between 9 and -9 T in Pr$_{0.5}$Ca$_{0.5}$Mn$_{0.95}$%
Ga$_{0.05}$O$_3$.

\item[FIG. 2.]  First magnetization curves at 5 K in four virgin Pr$_{0.5}$Ca%
$_{0.5}$Mn$_{0.95}$Ga$_{0.05}$O$_3$ samples coming from the same batch. The
arrows just mark the average values of the three step fields.

\item[FIG. 3.]  Magnetization loops in Pr$_{0.5}$Ca$_{0.5}$Mn$_{0.95}$Ga$%
_{0.05}$O$_3$ at various temperatures.

\item[FIG. 4.]  Five successive magnetization loops at 5 K in Pr$_{0.5}$Ca$%
_{0.5}$Mn$_{0.95}$Ga$_{0.05}$O$_3$. The labels refer to the run numbering.
Each loop was recorded after a zero-field cooling from room temperature.

\item[FIG. 5.]  High-field parts of hysteresis loops at 5 K in Pr$_{0.5}$Ca$%
_{0.5}$Mn$_{0.95}$Ga$_{0.05}$O$_3$. They were successively recorded using
different values of the field spacing ($f_s$) and of the waiting time ($t_w$%
) after each field installation (first loop: circles; second loop: squares;
third loop: diamonds).

\item[FIG. 6.]  Magnetization loops at 2.5 K in three ``pure-Mn''
manganites: a Pr$_{0.63}$Ca$_{0.37}$MnO$_3$ crystal, as well as Pr$_{0.65}$Ca%
$_{0.35}$MnO$_3$ and Pr$_{0.6}$Ca$_{0.4}$MnO$_3$ ceramics.
\end{description}


\begin{references}
\bibitem{REVIEW}  For a review, see {\it Colossal Magnetoresistance, Charge
Ordering and Related Properties of Manganese Oxides, }edited by C.\ N.\ R.\
Rao and B.\ Raveau (World Scientific, Singapore, 1998) and {\it Colossal
Magnetoresistance Oxides}, edited by Y.\ Tokura (Gordon \& Breach, London,
1999).

\bibitem{TOK98}  M. Tokunaga, N. Miura, Y. Tomioka, and Y. Tokura, Phys.
Rev. B. {\bf 57}, 5259 (1998); M. Respaud, A. Llobet, C. Frontera, C.
Ritter, J. M. Broto, H. Rakoto, M. Goiran, and J. L. Garc\'{i}a-Mu\~{n}oz, 
{\it ibid.} {\bf 61}, 9014 (2000).

\bibitem{LEE95}  M. R. Lees, J. Barratt, G. Balakrishnan, D. McK. Paul, and
M. Yethiraj, Phys. Rev. B. {\bf 52}, R14 303 (1995); Y. Tomioka, A.
Asamitsu, H. Kuwahara, Y. Morimoto, and Y. Tokura, {\it ibid.} {\bf 53},
R1689 (1996).

\bibitem{RAV97}  B. Raveau, A. Maignan and C. Martin, J. Solid State Chem. 
{\bf 130}, 162 (1997).

\bibitem{STEP1}  S. H\'{e}bert, V. Hardy, A. Maignan, R. Mahendiran, M.
Hervieu, C. Martin, and B. Raveau, J. Solid State Chem. {\bf 165}, 6 (2002).

\bibitem{STEP2}  S. H\'{e}bert, A. Maignan, V. Hardy, C. Martin, M. Hervieu,
and B. Raveau, submitted to Solid State Commun..

\bibitem{0604n1}  V. Hardy, A. Wahl, C. Martin, and Ch. Simon, Phys. Rev. B 
{\bf 63}, 224403 (2001); V. Hardy, A. Wahl, and C. Martin, {\it ibid.} {\bf %
64}, 064402 (2001).

\bibitem{Rq1}  In a Pr$_{0.5}$Ca$_{0.5}$Mn$_{1-x}$M$_x$O$_3$ compound, where
M$^{v+}$ is a nonmagnetic cation, the expected magnetization for full
polarization of the Mn spins is (in $\mu _B$ / f.u. ): $3.5+x\ (v-7)$ . Note
that the magnetization value at saturation can be slightly larger owing to
an additional contribution from the Pr spins.

\bibitem{STR77}  E. Stryjewski and N. Giordano, Adv. Phys. {\bf 26}, 487
(1977).

\bibitem{KUS96}  J. Kushauer, R. van Bentum, W. Kleemann, and D. Bertrand,
Phys. Rev. B {\bf 53}, 11 647 (1996).

\bibitem{MAR01}  C. Martin, A. Maignan, M. Hervieu, C. Autret, B. Raveau,
and D. I. Khomskii, Phys. Rev. B {\bf 63}, 174402 (2001).

\bibitem{UEH00}  M. Uehara and S.-W. Cheong, Europhys. Lett. {\bf 52}, 674
(2000).

\bibitem{MAH01}  R. Mahendiran, B. Raveau, M. Hervieu, C. Michel, and A.
Maignan, Phys. Rev. B {\bf 64}, 064424 (2001).

\bibitem{POD01}  V. Podzorov, B. G. Kim, V. Kiryukhin, M. E. Gershenson, and
S.-W. Cheong, Phys. Rev. B {\bf 64}, 140406(R) (2001).

\bibitem{VIV94a}  E. Vives, J. Ort\'{i}n, L. Ma\~{n}osa, I. R\'{a}fols, R.
P\'{e}rez-Magran\'{e}, and A. Planes, Phys. Rev. Lett. {\bf 72}, 1694 (1994).

\bibitem{DAM98}  F. Damay, C. Martin, A. Maignan, M. Hervieu, B. Raveau, G.
Andr\'{e}, and F. Bour\'{e}e, Appl. Phys. Lett. {\bf 73}, 3772 (1998).

\bibitem{JIR00}  Z. Jir\'{a}k, F. Damay, M. Hervieu, C. Martin, B. Raveau,
G. Andr\'{e}, and F. Bour\'{e}e, Phys. Rev. B {\bf 61}, 1181 (2000).

\bibitem{SKR91}  B. Skrotzki, J. Phys. IV {\bf 1}, c4-367 (1991).

\bibitem{HUM91}  J. Van Humbeeck, J. Phys. IV {\bf 1}, c4-189 (1991).

\bibitem{VIV94b}  E. Vives and A. Planes, Phys. Rev. B {\bf 50}, 3839
(1994); K. Dahmen and J. Sethna, {\it ibid.} {\bf 53}, 14 872 (1996).

\bibitem{LEV00}  P. Levy, F. Parisi, G. Polla, D. Vega, G. Leyva, H. Lanza,
R. S. Freitas and L. Ghivelder, Phys. Rev. B {\bf 62}, 6437 (2000).

\bibitem{SAK95}  H. Sakamoto, K. Meguro, A. Tanaka, and A. Imai, J. Phys. IV 
{\bf 5}, c8-581 (1995); M. Hayakawa, T. Okuda, and M. Oda, {\it ibid.} {\bf 5%
}, c8-1127 (1995).

\bibitem{PER01}  F. J. P\'{e}rez-Reche, E. Vives, L. Ma\~{n}osa, and A.
Planes, Phys. Rev. Lett. {\bf 87}, 195701 (2001).
\end{references}
\end{document}